\documentclass[aps,amsfonts,prl,twocolumn,superscriptaddress,showpacs]{revtex4}
\usepackage{epsfig,amsmath,amssymb,bm,epsf,graphics}

\def\ket#1{\vert#1\rangle}
\def\ketbra#1{\vert#1\rangle\langle#1\vert}
\def\ketbraS#1{\vert#1\rangle\!\langle#1\vert}
\def\ipr#1#2{\langle#1\vert#2\rangle}

\def\Longarrow{\protect\@lra}
\def\@lra{\relbar\joinrel\relbar\joinrel\relbar\joinrel%
          \relbar\joinrel\rightarrow}

\def\Chi{X}
\begin{document}
\title{Measures of entanglement in multipartite bound entangled states}

\author{Tzu-Chieh Wei}
\affiliation{Department of Physics, 
 University of Illinois at Urbana-Champaign, 
 1110 West Green Street, Urbana, Illinois 61801-3080, USA}
 \author{Joseph B. Altepeter}
\affiliation{Department of Physics, 
 University of Illinois at Urbana-Champaign, 
 1110 West Green Street, Urbana, Illinois 61801-3080, USA}
 \author{Paul M. Goldbart}
\affiliation{Department of Physics, 
 University of Illinois at Urbana-Champaign, 
 1110 West Green Street, Urbana, Illinois 61801-3080, USA}
\author{William J. Munro}
\affiliation{Hewlett-Packard Laboratories, Filton Road, Stoke Gifford,
         Bristol, BS34 SQ2, United Kingdom}
\date{\today}

\begin{abstract}
Bound entangled states are states that are entangled but
from which no entanglement can be distilled if all parties are
allowed only local operations and classical communication.
However, in creating these states one needs nonzero
entanglement resources to start with. Here, the entanglement of two distinct
multipartite bound entangled states is determined analytically in terms
of a geometric measure of entanglement and a related quantity.
The results are compared with those for the negativity and the relative entropy of entanglement. 

\end{abstract}
\maketitle

\noindent
{\it Introduction\/}---We are motivated to study the quantification of entanglement 
for the basic reason that entanglement 
has been identified as a {\it resource\/} central to much of quantum 
information processing (see, e.g., Ref.~\cite{NielsenChuang00}).  To date, progress 
in the quantification of entanglement for mixed states has resided 
primarily in the domain of bipartite systems~\cite{Horodecki01}.  
For multipartite systems in pure and mixed states, the characterization and quantification of entanglement present even greater challenges.  

Among all entangled states, there is a peculiar
class of states, bound entangled states, originally discovered
in the bipartite setting, where some connection to zero negativity was identified~\cite{Horodecki98}. These are
states that are entangled, but from which no pure entangled state
can be distilled if all parties are allowed only local operations
and classical communication (LOCC). The distillable entanglement 
($E_{\rm D}$) is thus zero. Bound entangled states can be either
bipartite or multipartite, the latter possibly exhibiting more
structure than the former. However, it does take nonzero 
entanglement to {\it create\/} bound entangled states under LOCC. It is thus desirable to see how
much entanglement  is needed. 

However, the two most important 
measures---entanglement of 
distillation ($E_{\rm D}$) and of formation ($E_{\rm F}$)---have so far been limited to {\it bipartite\/} settings,
as there are ambiguities in generalizing them 
to multipartite systems~\cite{PlenioVedral01}.
In order to explore {\it multipartite\/} settings, it is thus, on the one hand,
necessary to lay down bounds on the entanglement content for distillation
and formation. On the other hand, applying other measures,
such as the relative entropy of entanglement ($E_{\rm R}$),
to multipartite states may prove helpful in quantifying entanglement.
Recently, a multipartite entanglement measure based on the geometry
of Hilbert space has been proposed~\cite{Shimony95,BarnumLinden01,WeiGoldbart02}, 
and has been applied to several bipartite and multipartite cases.
The merit of this measure is that it is suitable for any-partite systems
with any dimension, although determining it analytically for
generic states remains a challenge.

In the present paper, we study the entanglement content of
two distinct bound entangled states:
Smolin's four-party unlockable bound entangled state~\cite{Smolin00,ShorSmolinThapliyal03} and 
D\"ur's $N$-party Bell-inequality-violating bound entangled states~\cite{Dur01}. For each, we  
determine analytically  their geometric measure of entanglement
and a related quantity. 
Under certain circumstances, these give lower bounds on their multipartite $E_{\rm F}$. 
In addition, we make conjectures concerning their relative entropies of 
entanglement.  Although quantities such as the geometric
measure or the relative entropy of entanglement may not be able to
reveal the exact nature of bound entanglement, they nevertheless quantify
for these bound entangled states the content of entanglement that is unextractable. 

\smallskip
\noindent
{\it Geometric measure of entanglement}---We begin by  briefly reviewing its formulation. 
Consider a general $n$-partite pure state (expanded in the local bases 
$\{e_{p_i}^{(i)}\}$): 
\begin{equation}
|\psi\rangle=\sum_{p_1\cdots p_n}\chi_{p_1p_2\cdots p_n}
|e_{p_1}^{(1)}e_{p_2}^{(2)}\cdots e_{p_n}^{(n)}\rangle.
\end{equation}
As shown in Ref.~\cite{WeiGoldbart02}, its closest {\it separable} (i.e., product) pure state (with $i$ being the party index)
\begin{equation}
\ket{\phi}\equiv\mathop{\otimes}_{i=1}^n|\phi^{(i)}\rangle=\mathop{\otimes}_{i=1}^{n}
\Big({\sum}_{p_i}c_{p_i}^{(i)}\,|e_{p_i}^{(i)}\rangle\Big),
\end{equation}
satisfies the condition (and its complex conjugate)
\begin{equation}
\sum_{p_1\cdots\widehat{p_i}\cdots p_n}
\chi_{p_1p_2\cdots p_n}^*c_{p_1}^{(1)}\cdots\widehat{c_{p_i}^{(i)}}\cdots c_{p_n}^{(n)}=
\Lambda\,{c_{p_i}^{(i)}}^*, 
\end{equation}
where the eigenvalue $\Lambda\in[-1,1]$ is associated with the Lagrange 
multiplier enforcing  
$\ipr{\phi}{\phi}\!=\!1$, 
and the caret denotes exclusion.  
Moreover, $\Lambda$ is the cosine of the
angle between $|\psi\rangle$ and $\ket{\phi}$; 
the largest one $\Lambda_{\max}$, which we call the 
{\it entanglement eigenvalue\/}, corresponds to the closest 
separable state, and is the maximal overlap: 
$\Lambda_{\max}(\ket{\psi})=\max_{\phi}|\ipr{\phi}{\psi}|$,
where $\ket{\phi}$ is separable but otherwise arbitrary. 
$E_{\sin^2}\equiv 1-\Lambda^2_{\max}(\ket{\psi})$ was defined to be the
geometric measure of entanglement~\cite{WeiGoldbart02} for  state $|\psi\rangle$, and it measures the degree of inseparability via
the squared sine of the angle away from the closest separable pure state. 

The extension  to mixed states  can be built upon  the pure-state theory
and is made via the use of the 
{\it convex hull\/} construction, 
as was done for $E_{\rm F}$~\cite{Horodecki01}.  The essence is a minimization 
over all decompositions $\rho=\sum_i p_i\,|\psi_i\rangle\langle\psi_i|$ 
into pure states: 
\begin{eqnarray}
\label{eqn:Emixed}
E_{\sin^2}(\rho)
\equiv
{\min_{\{p_i,\psi_i\}}}
\sum\nolimits_i p_i \, 
E_{\sin^2}(|\psi_i\rangle).
\end{eqnarray}
This convex hull construction ensures that the measure gives zero 
for unentangled states; however, it also complicates the 
task of determining mixed-state entanglement.  
$E_{\sin^2}$ was shown to be
 an {\it entanglement
monotone\/}~\cite{BarnumLinden01,WeiGoldbart02} (i.e., the measure does not increase under LOCC), hence is a good measure of entanglement.
As there is no explicit generalization of $E_{\rm F}$ to multipartite states,
we shall calculate  $E_{\sin^2}$ analytically 
for two bound entangled states, Smolin's and D\"ur's.
Because $E_{\rm F}(\rho)$ is the minimum average {\it ebit\/} to 
create a single copy of $\rho$, we can regard $E_{\sin^2}(\rho)$ as
the minimum average degree of pure inseparability needed to realize the state $\rho$.

In bipartite settings, it is known~\cite{VedralPlenio98}
that $E_{\rm R}(\rho)\le E_{\rm F}(\rho)$, and that
for pure states $\ket{\psi}$, $E_{\rm R}({\psi})=E_{\rm F}({\psi})$.
It is also known~\cite{WeiErricsonGoldbartMunro} that for
any bi- and multipartite {\it pure\/} state
$\ket{\psi}$, ${\cal E}_{\log_2}(\psi)\equiv- 2\log_2\Lambda_{\max}(\psi) \le E_{\rm R}(\psi)$.
Together with the inequality $(1-x^2)\log_2e\le -2\log_2 x$  (for $0\le x\le 1$),
one has
\begin{eqnarray}
\sum\nolimits_i\! p_i \, 
E_{\sin^2}(\psi_i)\log_2e\le\!
\sum\nolimits_i\! p_i \, 
{\cal E}_{\log_2}(\psi_i)\le 
\!\sum\nolimits_i p_i \, 
{E}_{\rm F}(\psi_i), \nonumber
\end{eqnarray}
and thus $(\log_2e)E_{\sin^2}(\rho)\le {\cal E}_{\log_2}(\rho)\le E_{\rm F}(\rho)$
for any {\it bipartite\/} state $\rho$. If the generalization of $E_{\rm F}$ to
multipartite states maintains the property that $E_{\rm F}(\psi)\ge E_{\rm
R}(\psi)$ then the inequality $(\log_2e)E_{\sin^2}\le {\cal E}_{\log_2} \le E_{\rm F}$
will continue to hold for multipartite mixed states.
We remark that 
\begin{equation}
{\cal E}_{\log_2}(\rho)\equiv 
{\min_{\{p_i,\psi_i\}}}\sum\nolimits_i p_i  
\Big(-2 \log_2\Lambda_{\max}(\psi_i)\Big)
\end{equation}
is not an entanglement monotone~\cite{WeiErricsonGoldbartMunro}. However, 
we see that both $(\log_2e)E_{\sin^2}$ and
${\cal E}_{\log_2}$ could serve as lower bounds on multipartite
entanglement of formation.

We now turn to the calculations of entanglement for the two bound entangled states: Smolin's and D\"ur's.

\smallskip
\noindent
{\it Smolin's four-party unlockable bound entangled state\/}---This is a four-qubit mixed state
\begin{equation}
\label{eqn:rhoABCD}
\rho^{ABCD}\equiv\frac{1}{4}\sum_{i=0}^3\big(\ketbra{\Psi_i}\big)_{\rm AB}\otimes
\big(\ketbra{\Psi_i}\big)_{\rm CD},
\end{equation}
where the $\ket{\Psi}$'s are the four Bell states $(\ket{00}\pm\ket{11})/\sqrt{2}$
and $(\ket{01}\pm\ket{10})/\sqrt{2}$.
Now, the state $\rho^{ABCD}$ can be conveniently rewritten as 
\begin{equation}
\label{eqn:rhoABCD2}
\rho^{ABCD}=\frac{1}{4}\sum_{i=0}^3\ketbra{\Chi_i},
\end{equation}
where the $\ket{\Chi}$'s are the 
four orthogonal Greenberger-Horne-Zeilinger-(GHZ) like states:
\begin{eqnarray}
&&\!\!\!\!\!\!
\ket{\Chi_0}\!\equiv\!\frac{1}{\sqrt{2}}\big(\ket{0000}\!+\!\ket{1111}\big),
\
\ket{\Chi_1}\!\equiv\!\frac{1}{\sqrt{2}}\big(\ket{0011}\!+\!\ket{1100}\big),\nonumber\\
&&\!\!\!\!\!\!
\ket{\Chi_2}\!\equiv\!\frac{1}{\sqrt{2}}\big(\ket{0101}\!+\!\ket{1010}\big), 
\
\ket{\Chi_3}\!\equiv\!\frac{1}{\sqrt{2}}\big(\ket{0110}\!+\!\ket{1001}\big).\nonumber
\end{eqnarray}

The most general decomposition of a mixed state $\rho$ into pure states
can be expressed as
\begin{equation}
\label{eqn:rhoDecomp}
\rho=\sum_{k=1}^{\cal M} 
\ketbra{\tilde{\varphi}_k}\ \ {\rm with} \
\ket{\tilde{\varphi}_k}=\sum_{i=1}^n {\cal U}_{ki}\sqrt{\lambda_i}\,\ket{\xi_i},
\end{equation}
where ${\cal M}$ is an integer not smaller than $n$, the
number of orthonormal eigenvectors $\{\ket{\xi_i}\}$ (with nonzero eigenvalues $\{\lambda_i\}$) of $\rho$, the  
$\ket{\tilde{\varphi}}$'s are {\it unnormalized\/}, and 
${\cal U}$ satisfies
$\sum_{k=1}^{\cal M}{\cal U}_{ki}\,{\cal U}^*_{kj}=\delta_{ij}$.
Thus, the most general pure state that appears in the decomposition of Smolin's state is
\begin{equation}
\label{eqn:varphi}
\ket{\tilde{\varphi}_k}=\sum_{i=0}^3 \frac{1}{2}\,{\cal U}_{ki}\ket{\Chi_i}.
\end{equation}
Our goal is to minimize
$\sum_{k} p_k \, E_{\rm pure}\big(\ket{{\varphi}_k}\big)$
over all possible ${\cal U}$'s, where $E_{\rm pure}$ is some pure-state
entanglement ($E_{\sin^2}$ or ${\cal E}_{\log_2}$ in our
considerations), $p_k\equiv \ipr{\tilde{\varphi}_k}{\tilde{\varphi}_k}$,
and $\ket{\varphi_k}$ is the normalized state $\ket{\varphi_k}\equiv \ket{\tilde{\varphi}_k}/\sqrt{p_k}$. Making a general minimization for an arbitrary 
mixed state is extremely difficult. However, for the mixed state $\rho^{ABCD}$
we shall show that the decomposition in
Eq.~(\ref{eqn:rhoABCD2}) does indeed minimize the average entanglement
over pure-state decompositions. As in Eq.~(\ref{eqn:varphi}), $\ket{\varphi}$ 
can be explicitly written as 
$\ket{{\varphi}}=\sum_{i=0}^3 \sqrt{q_i}\, e^{i\,\phi_i}\ket{\Chi_i}$, 
where the $q$'s are non-negative, satisfying $\sum_iq_i=1$, and
the $\phi$'s are phases.
For fixed $q$'s, the state has a maximal
entanglement eigenvalue  when all phases
are zero. We shall show shortly that its maximal entanglement
eigenvalue is $1/\sqrt{2}$,
which is achieved by the $\ket{\Chi}$'s.

The entanglement eigenvalue of the state
$\ket{{\varphi}}=\sum_{i=0}^3 \sqrt{q_i}\, \ket{\Chi_i}$
is the maximal overlap with the separable state 
$\ket{\Phi}=\otimes_{i=1}^4 \big(c_i\ket{0}+s_i\ket{1}\big)$,
where $c_i\equiv\cos\theta_i$ and $s_i\equiv\sin\theta_i$
with $0\le \theta_i\le \pi/2$. Thus 
\begin{eqnarray}
&&\!\!\!\!\!\!\!\!\ipr{\Phi}{{\varphi}}=
\sqrt{{q_0}/{2}}\,(c_1c_2c_3c_4
                    +s_1s_2s_3s_4)+		    \sqrt{{q_1}/{2}}\,(c_1c_2s_3s_4\nonumber \\
                    && \qquad+s_1s_2c_3c_4)\,
+		    \sqrt{{q_2}/{2}}\,(c_1s_2c_3s_4
                    +s_1c_2s_3c_4)\nonumber\\
&&		   \qquad +\sqrt{{q_3}/{2}}\,(c_1s_2s_3c_4
                   +s_1c_2c_3s_4),
\nonumber
\end{eqnarray}
which has maximum $1/\sqrt{2}$. To see this, use the
Cauchy-Schwarz inequality, treating as one
vector
$\big\{\sqrt{q_0/2},\sqrt{q_1/2},\sqrt{q_2/2},\sqrt{q_3/2}\big\}$ (whose
modulus is $1/\sqrt{2}$\,), and the corresponding coefficients as another vector (whose
modulus can be shown to be no greater than 1; see App. A). The states $\ket{\Chi_i}$ clearly saturate this bound;
hence:
\begin{equation}
E_{\sin^2}(\rho^{ABCD})={1}/{2} , \ \
{\cal E}_{\log_2}(\rho^{ABCD}) = 1.
\end{equation}
This suggests that although bound entangled, 
Smolin's state has a very high degree of entanglement, the same as that
of a 4-partite GHZ state. This high degree of entanglement seems to manifest
in some bipartite partitioning, e.g., \{A:BCD\} (as we discuss below).

We conjecture (and later prove) that $E_{\rm R}\!=\!1$ for this state
and one of its closest
separable mixed states is 
\begin{eqnarray}
&&\!\!\!\!\!\!\!\!\frac{1}{8}\!\big(\ketbraS{0000}\!+\!\ketbraS{1111}\!+\!\ketbraS{0011}
\!+\!\ketbraS{1100}\nonumber\\
&&\!\!\!\!\!\!\!\!+\ketbraS{0101}\!+\!\ketbraS{1010}\!+\!\ketbraS{0110}\!+\!\ketbraS{1001}\big).\nonumber
\end{eqnarray} 
We remark that the negativity ${\cal N}$~\cite{Horodecki01} (a value used to
quantify the degree of bipartite inseparability of states and defined as twice the absolute sum
of negative eigenvalues of the partial transpose (PT) of the density matrix
with respect to some bipartite partitioning) is zero for any 2/2 partitioning, e.g., \{AB : CD\}, but nonzero for 1/3 partitioning, e.g.,\{A:BCD\}.  Specifically, ${\cal N}_{\rm A:BCD}=1$
but ${\cal N}_{\rm AB:CD}=0$.

\smallskip
Let us now turn to D\"ur's bound entangled states.

\noindent
{\it D\"ur's $N$-party  bound entangled states\/}---D\"ur~\cite{Dur01} found that for $N\ge 4$ the following state is bound
entangled:
\begin{equation}
\rho_N\equiv\frac{1}{N+1}\left(\ketbra{\Psi_G}+\frac{1}{2}\sum_{k=1}^N\big(P_k+\bar{P}_k\big)\right),
\end{equation}
where $\ket{\Psi_G}\equiv 
\big(\ket{0^{\otimes N}}+e^{i\alpha_N}\ket{1^{\otimes N}}\big)/        {\sqrt{2}}$ is a $N$-partite GHZ state;
$P_k\equiv\ketbra{u_k}$ is a projector onto the state $\ket{u_k}\equiv\ket{0}_1\ket{0}_2\ldots\ket{1}_k\ldots\ket{0}_N$;
and
$\bar{P}_k\equiv\ketbra{v_k}$ projects on to $\ket{v_k}\equiv\ket{1}_1\ket{1}_2\ldots\ket{0}_k\ldots\ket{1}_N$.
For $N\ge 8$ this state violates
the Mermin-Klyshko-Bell inequality~\cite{Dur01}; 
violation was pushed down to $N\ge 7$ by Kaszlikowski et al.~\cite{Kwek02} for a three-setting
Bell inequality; it was pushed further down to
$N\ge 6$ by Sen {\it et al.\/}~\cite{SenSenZukowski02} for
a functional Bell inequality.
The phase $\alpha_N$ in
$\ket{\Psi_G}$ can be eliminated by local unitary transformations, and hence we shall take $\alpha_N=0$ in
the following discussion.

In fact, if we consider the family of states
\begin{equation}
\rho_N(x)\equiv x\ketbra{\Psi_G}+\frac{1-x}{2N}
\sum_{k=1}^N\big(P_k+\bar{P}_k\big),
\end{equation}
we find that
for $N\ge 4$ the state is bound entangled if $0<x\le 1/(N+1)$ and is still entangled but not bound entangled
if $x > 1/(N+1)$. This can be seen from 
the fact that the negativities of $\rho_N(x)$ 
with respect to the two different partitions
$(1:2\cdots N)$ and $(12:3\cdots N)$ are
\begin{subequations}
\begin{eqnarray}
&&\!\!\!\!\!\!\!\!\!\!{\cal N}_{1:2\cdots
 N}\big(\rho_N(x)\big)=\max\left\{0,[{(N\!+\!1)\,x-1}\,]/{N}\,\right\}, \\
 &&\!\!\!\!\!\!\!\!\!\!{\cal N}_{12:3\cdots
 N}\big(\rho_N(x)\big)=x.
\end{eqnarray}
\end{subequations}

By applying arguments similar to those used to calculate entanglement
for Smolin's state, we have that the general pure state
in the decomposition of $\rho_N(x)$ is
\begin{equation}
\sqrt{y}\,e^{i\phi_0}\ket{\Psi_G}+\sqrt{1\!-\!y}\sum_{k=1}^N
\big(\sqrt{q_k}e^{i \phi_{i}}\ket{u_i}+
\sqrt{r_k}e^{i \phi'_{i}}\ket{v_i}\big),\nonumber
\end{equation}
where the $q$'s and $r$'s are non-negative and satisfy $\sum_k(q_k+r_k)=1$.
In this family, the state with the least entanglement (or
maximum $\Lambda_{\max}$) for fixed $\{y,q_k,r_k\}$ is the
one with all phase factors zero:
\begin{equation}
\ket{\Psi\big(y,\{q,r\}\big)}\equiv\sqrt{y}\ket{\Psi_G}+\sqrt{1\!-\!y}\sum_{k=1}^N
\big(\sqrt{q_k}\ket{u_i}+
\sqrt{r_k}\ket{v_i}\big).\nonumber
\end{equation}
Next, we ask: For fixed $y$, what is the least entanglement
that the above state can have?
Take a separable state of the form
$\ket{\Phi}=\otimes_{i=1}^N
\big(c_i\ket{0}+s_i\ket{1}\big)$; 
its overlap with $\ket{\Psi\big(y,\{q,r\}\big)}$ is then
\begin{eqnarray}
&&\ipr{\Psi}{\Phi}
=\sqrt{{y}/{2}}\,
(c_1\cdots c_N+s_1\cdots s_N)
\nonumber \\
&&+\sqrt{1\!-\!y}\sum_{k=1}^N
(\sqrt{q_k}\,c_1\cdots s_k\cdots c_N
\!+\!\sqrt{r_k}\,s_1\cdots c_k\cdots s_N).\nonumber
\end{eqnarray}
This can be shown to be no greater than $\sqrt{(2-y)/2}$,
again by a Cauchy-Schwarz inequality, taking
\begin{equation}
\left\{ \sqrt{{y}/{2}}, \big\{\sqrt{(1-y)q_k}\big\},
 \big\{\sqrt{(1-y)r_k}\big\}\right\}\nonumber
\end{equation} 
as the first $(2N\!+\!1)$-component vector (with modulus $\sqrt{(2\!-\!y)/2}$) and the
corresponding coefficients as the second one (whose modulus can
be shown to be no greater than 1 for $N\ge 4$; see App. A).
The bound can be saturated, e.g., by  
\begin{subequations}
\label{eqn:psiuv}
\begin{eqnarray}
&&\ket{\psi_{\pm,u,k}(y)}\equiv\sqrt{y}\ket{\Psi_G}\pm\sqrt{1-y}\ket{u_k},\\
&&\ket{\psi_{\pm,v,k}(y)}\equiv\sqrt{y}\ket{\Psi_G}\pm\sqrt{1-y}\ket{v_k},
\end{eqnarray}
\end{subequations}
for which $\Lambda_{\max}(y)=\sqrt{(2-y)/2}$~\cite{footnote}.
As $1-\Lambda^2_{\max}(y)$ is linear in $y$ and $-2\log_2\Lambda_{\max}(y)$
is convex in $y$, one gets
\begin{equation}
E_{\sin^2}(\rho_N(x)) =\frac{x}{2}, \ \
{\cal E}_{\log_2}(\rho_N(x))=\log_2\frac{2}{2-x},
\end{equation}
and one of the optimal decompositions is  
\begin{equation}
\rho_N(x)=\frac{1}{4N}\sum_{k=1}^{N}\sum_{\alpha=\pm}\sum_{\beta=u,v}\ketbra{\psi_{\alpha,\beta,k}(x)}.
\end{equation}
The above calculations show that for $\rho_N(x)$, the entanglement
depends on the portion $x$ of the GHZ state in states $\ketbra{\psi_{\alpha,\beta,k}(x)}$ and it never becomes zero
unless there is no GHZ mixture. 

We conjecture that, for $N\ge 4$, $\rho_N(x)$ has $E_R(x)=x$, with
one closest separable mixed state being 
\begin{equation}
\frac{x}{2}\big(\ketbra{0..0}+\ketbra{1..1}\big)
+\frac{1-x}{2N}\sum_{k=1}^N\big(P_k +\bar{P}_k\big),\nonumber
\end{equation}
which seems plausible as $\big(\ketbra{0..0}+\ketbra{1..1}\big)$ 
is a closest
separable mixed state to  $\ket{\Psi_G}$. 

\smallskip
\noindent
{\it Concluding remarks\/}---We have presented analytical results on how much entanglement is bound in two distinct multipartite bound entangled states.
The measure we have used to quantify their entanglement
is the geometric measure of entanglement (GME), whose construction, similiar
to the entanglement of formation ($E_{\rm F}$), is via the convex hull. In contrast
to GME, $E_{\rm F}$ has not been explicitly generalized to multipartite states,
and hence is still unavailable for these bound entangled states. However, under
the circumstances discussed previously, the results for $E_{\sin^2}$ as well as a related quantity,
${\cal E}_{\log_2}$, might provide lower bounds on $E_{\rm F}$.
 For the Smolin state, its bound entanglement is
as large as that of a four-partite GHZ state, whereas that
for D\"ur states is related to the portion of the $N$-partite
GHZ state. For each case, an optimal decomposition is given. Furthermore, we have conjectured that the relative
entropy of entanglement ($E_{\rm R}$) for the Smolin state is unity (proved below), whereas we
conjecture that $E_{\rm R}$ for D\"ur's state is equal to
the portion that is $N$-partite GHZ. 

For  Smolin's  state
we can establish its $E_{\rm F}$, $E_{\rm D}$, $E_{\rm R}$,
and $E_{\sin^2}$ 
for certain bipartite
partitionings. For example,
if we group the four parties ABCD in two, A:BCD, we can write the state as
\begin{equation}
\rho^{A:BCD}=\frac{1}{4}\sum_{i=0}^3\ketbra{\bar{\Chi}_i},
\end{equation}
with the 3-qubit states of BCD mapped on to the 8-level system ($000\rightarrow \underline{0}, 001\rightarrow\underline{1},..., 111\rightarrow\underline{7}$), 
involving the locally orthogonal and convertible
states (by BCD)
\begin{eqnarray}
&&\!\!\!\!\!
\ket{\bar{\Chi}_0}=\big(\ket{0\underline{0}}\!+\!\ket{1\underline{7}}\big)/{\sqrt{2}},
\ \ \
\ket{\bar{\Chi}_1}=\big(\ket{0\underline{3}}\!+\!\ket{1\underline{4}}\big)/{\sqrt{2}},\nonumber\\
&&\!\!\!\!\!
\ket{\bar{\Chi}_2}=\big(\ket{0\underline{5}}\!+\!\ket{1\underline{2}}\big)/{\sqrt{2}},
\ \ \
\ket{\bar{\Chi}_3}=\big(\ket{0\underline{6}}\!+\!\ket{1\underline{1}}\big)/{\sqrt{2}}.\nonumber
\end{eqnarray}
In order to find the entanglement of this bipartite state (in $C^{2}\otimes C^{8}$),
we need to consider the entanglement of the general  (properly
normalized) pure state 
\begin{equation}
\ket{\psi}\equiv{\sum}_i\sqrt{x_i}\, e^{i\phi_i}\ket{\bar{\Chi}_i} \nonumber
\end{equation}
that appears
in the pure-state decompositions. In fact, regardless of the
values of the ${x_i}$'s, this pure state  has
a reduced density matrix (tracing over BCD) of the form 
$\left(\ketbra{0}+\ketbra{1}\right)/2$.
This shows that $\rho^{A:BCD}$ has 
$E_{\rm F}=1$, $E_{\sin^2}=1/2$, and ${\cal E}_{\log_2}=1$.
In fact, there is a general result due to Horodecki et al.~\cite{Horodecki398a}
that $E_{\rm D}=E_{\rm F}$ for mixtures of locally orthogonal bipartite states, e.g., 
$C^2\otimes C^{2m}$ states
that are derived from mixing Bell-like states 
\begin{equation}
\label{eqn:Belllike}
\ket{\Psi^\pm_k}\equiv(\ket{0,\underline{k}}\pm\ket{1,\underline{2m-k-1}})/{\sqrt{2}},
\end{equation}
having {\it distinct\/} $k$'s, where $k=0,1,\ldots, m-1$. 
 As $E_{\rm D}\le E_{\rm R}\le E_{\rm F}$, we have that $E_{\rm R}(\rho^{\rm A:BCD})=1$ as well. What about the original four-partite state
$\rho^{\rm ABCD}$? As $E_{\rm R}(\rho^{\rm ABCD})\ge E_{\rm R}(\rho^{\rm A:BCD})$, we
have $E_{\rm R}(\rho^{\rm ABCD})\ge 1$. But we also have that $E_{\rm R}(\rho^{\rm ABCD})\le 1$, as our previous conjecture gives at least
an upper bound; we thus have that $E_{\rm R}(\rho^{\rm ABCD})=1$ and
the conjecture is proved. Naively, we expect that any arbitrary
$\rho^{\rm ABCD}$ has greater entanglement than $\rho^{\rm A:BCD}$.
However, for the Smolin state, they have the same entanglement
as quantified by both GME and the relative entropy of entanglement.

Although D\"ur's bound entangled state violates a Bell inequality,
it has nonzero negativity under certain partitionings. One may
raise the question: Does there exist a bound entangled state
that has positive PT (PPT) under all partitionings but that  still violates
a Bell's inequality? For example, does an unextendible-product-basis (UPB) bound entangled state~\cite{BennettDiVincenzoMorShorSmolinTerhal99}
violate a Bell inequality? We shall see shortly that the answer
is ``no'', at least for the three different Bell
inequalities~\cite{Dur01,Kwek02,SenSenZukowski02} mentioned
earlier. Ac\'\i n has shown~\cite{Acin02}
that if an $N$-qubit state violates a two-setting Bell inequality then 
it is distillable under certain  bipartite
partitioning. Using the results of Refs.~\cite{DurCirac00,DurCirac00B} regarding
distillability, we can repeat the same analysis for the other
two  inequalities~\cite{Kwek02,SenSenZukowski02} and indeed obtain
the same conclusion; see App. B. This bipartite distillability then implies a 
negative PT (NPT) 
under that bipartite partitioning according to Horodecki et al.~\cite{Horodecki98}
Hence, violating
these Bell inequalities implies NPT under certain bipartite partitioning.
Said equivalently, if an $N$-qubit state has PPT under 
all bipartite partitionings then
the state never violates these Bell inequalities. This seems to
suggest that PPT bound entangled states are truly bound in nature
that cannot give deviation from local theories.

\smallskip
\noindent
{\it Acknowledgments\/}: 
We thank M.~Ericsson, P.~Kwiat,
S.~Mukhopadhyay, O. Rudolph and especially W.~D\"ur for discussions.
This work was supported by 
NSF Grant No. EIA01-21568
and DOE Grant No. DEFG02-91ER45439.

\smallskip
\noindent {\it Appendix A:\/}
In this appendix we sketch proofs of two useful inequalities and
describe the
deriviation of the entanglement eigenvalue for the states in Eqs.~(\ref{eqn:psiuv}).
We start with 
the first sought inequality: 
\begin{eqnarray}
&&(c_1c_2c_3c_4
                 +s_1s_2s_3s_4)^2
+		    (c_1c_2s_3s_4
                    +s_1s_2c_3c_4)^2 \nonumber \\
&&+		    (c_1s_2c_3s_4
                    +s_1c_2s_3s_4)^2
+		    (c_1s_2s_3c_4
                    +s_1c_2c_3s_4)^2\le 1\nonumber.
\end{eqnarray}
We have simplified the notation by using $c_i\equiv \cos\theta_i$ and
$s_i\equiv \sin\theta_i$. By subtracting the left-hand side from 1
and doing some algebraic manipulation,  we arrive at the non-negative
expression (hence the sought result):
\begin{eqnarray}
 &&(c_1c_2c_3s_4-s_1s_2s_3c_4)^2+(c_1c_2s_3s_4-s_1s_2c_3s_4)^2+\nonumber \\ &&(c_1s_2c_3c_4-
 s_1c_2s_3s_4)^2+(s_1c_2c_3c_4-c_1s_2s_3s_4)^2\nonumber.
\end{eqnarray}
The next sought inequality is (for $N\ge 4$):
\begin{eqnarray}
&& f_N\equiv\big(c_1\cdots c_N+s_1\cdots s_N\big)^2
+\nonumber \\
&&\quad\sum_{k=1}^N\left\{
(c_1\cdots s_k\cdots c_N)^2
+(s_1\cdots c_k\cdots s_N)^2\right\}\le 1.\nonumber
\end{eqnarray}
First, making similar arguments, one
can show that $f_4\le 1$. One can also show that $f_{N+1}\le f_N$.
Thus, by induction, we have the sought result.

We now discuss why  $\sqrt{y}\ket{\Psi_G}\pm\sqrt{1-y}\ket{u_k}$ and
$\sqrt{y}\ket{\Psi_G}\pm\sqrt{1-y}\ket{v_k}$
have as their  maximal entanglement eigenvalue 
 $\Lambda_{\max}(y)=\sqrt{(2-y)/2}$.
As one can make local relative phase shifts to transform
$\sqrt{y}\ket{\Psi_G}+\sqrt{1-y}\ket{u_k}$ to $\sqrt{y}\ket{\Psi_G}-\sqrt{1-y}\ket{u_k}$, they have the same entanglement. 
The change from $\sqrt{y}\ket{\Psi_G}\pm\sqrt{1-y}\ket{u_k}$ to
$\sqrt{y}\ket{\Psi_G}\pm\sqrt{1-y}\ket{v_k}$ is simply a flipping of 0 to 1, and
vice versa. The mapping from $k$ to $k'$ is just a relabeling of parties.
Thus, we need only consider the state
\begin{equation}
\sqrt{y/{2}}\,(\ket{00\cdots0}+\ket{11\cdots1})+\sqrt{1-y}
\ket{10\cdots0}.\nonumber
\end{equation}
As this state is invariant under permutation of all
parties except the first one, and as the coefficients are non-negative,
in order to find the maximal overlap
we can make the hypothesis that the closest separable state is of the form
\begin{equation}
\left(\sqrt{p}\ket{0}+\sqrt{1-p}\ket{1}\right)\otimes (\sqrt{q}\ket{0}+\sqrt{1-q}\ket{1})^{\otimes N-1}. \nonumber
\end{equation}
We further see that in order for the overlap to be maximal, $q$ must be either 1 or 0. For the former case,
we can further maximize the overlap to get $\sqrt{(2-y)/2}$. For
the latter case, the maximum overlap is $\sqrt{y/2}$, which is less
than $\sqrt{(2-y)/2}$ (as $0\le y\le 1$). Hence, the state  $\sqrt{y}\ket{\Psi_G}\pm\sqrt{1-y}\ket{u_k}$ has the entanglement eigenvalue
$\sqrt{(2-y)/2}$.

\smallskip
\noindent {\it Appendix B:\/}
In this appendix we analyze the connection between violation of three Bell inequalities and bipartite
distillability as was done in Ref.~\cite{Acin02} for the two-setting inequality. It was shown by D\"ur and Cirac~\cite{DurCirac00B} that an arbitrary $N$-qubit state $\rho$ can be locally depolarized
into the form
\begin{eqnarray}
\rho_N& =&\lambda_0^+\ketbra{\Psi^+_0}+\lambda_0^-\ketbra{\Psi^-_0}\nonumber \\
&+& \sum_{j=1}^{2^{N\!-\!1}-1}\lambda_j\big(\ketbra{\Psi^+_j}+\ketbra{\Psi^-_j}\big),\nonumber
\end{eqnarray}
while preserving $\lambda_0^\pm=\langle\Psi_0^\pm|\rho|\Psi_0^\pm\rangle$
and $\lambda_j=\langle\Psi_j^+|\rho|\Psi_j^+\rangle+\langle\Psi_j^-|\rho|\Psi_j^-\rangle$, where $\ket{\Psi^\pm_0}\equiv (\ket{0^{\otimes N}}\pm\ket{1^{\otimes N}})/\sqrt{2}$,
and the $\ket{\Psi^\pm_j}$'s are GHZ-like states, i.e., the states in Eq.~(\ref{eqn:Belllike}), unfolded into qubit notation. Normalization gives the condition
\begin{equation}
\lambda_0^++\lambda_0^-+2\sum_j\lambda_j=1.\nonumber
\end{equation}
Now define $\Delta\equiv \lambda_0^+-\lambda_0^-$, which we assume 
to be non-negative without loss of generality. 
The condition that there is no bipartite distillability for some bipartite
partitioning $P_j$ is~\cite{DurCirac00} 
\begin{equation}
2\lambda_j\ge\Delta.\nonumber
\end{equation}
Assuming nondistillability for {\it all\/} bipartite splittings, we have
\begin{equation}
2\sum_j \lambda_j = 1-(\lambda_0^++\lambda_0^-) \ge (2^{N\!-\!1}-1)\Delta.\nonumber
\end{equation}
As $\lambda_0^++\lambda_0^-\ge \Delta$, we have further that
\begin{equation}
\label{eqn:NoDistill}
1-\Delta \ge (2^{N\!-\!1}-1)\Delta.
\end{equation}
For the two-setting Bell inequality considered by Ac\'in~\cite{Acin02}, violation implies 
$\Delta> 1/ 2^{(N\!-\!1)/2}$. 
For the three-setting Bell inequality considered in~\cite{Kwek02},
violation 
implies
$\Delta> \sqrt{3} \,(2^N/3^N).$ 
For the functional Bell inequality in~\cite{SenSenZukowski02},
violation 
implies
$\Delta >2 \,(2^N/\pi^N).$ 
One can easily check that the three Bell inequalities considered are inconsistent
with the non-bipartite-distillability condition, Eq.~(\ref{eqn:NoDistill}).
Hence, the violating of these three Bell inequalities implies the existence
of some bipartite distillability.



\begin{thebibliography}{99}  
\bibitem{NielsenChuang00} 
M. Nielsen and I. Chuang,
{\sl Quantum Computation and Quantum Information\/}
(Cambridge Univ. Press, 2000).
\bibitem{Horodecki01}
For a review, see 
M. Horodecki, 
Quant. Info. Comp. {\bf 1\/}, 3 (2001), 
and references therein.
\bibitem{Horodecki98}
M. Horodecki, P. Horodecki, and R. Horodecki, Phys. Rev. Lett. {\bf 80}, 5239 (1998).
\bibitem{PlenioVedral01}
M. B. Plenio and V. Vedral, J. Phys. A {\bf 34}, 6997 (2001).
\bibitem{Shimony95}
A. Shimony, 
Ann. NY. Acad. Sci. {\bf 755\/}, 675 (1995).
\bibitem{BarnumLinden01}
H. Barnum and N. Linden,
J. Phys. A {\bf 34\/}, 6787 (2001).
\bibitem{WeiGoldbart02}
T.-C. Wei and P. M. Goldbart, Phys. Rev. A {\bf 68}, 042307 (2003).
\bibitem{Smolin00}
J. A. Smolin, 
Phys. Rev. A {\bf 63}, 032306 (2001).
\bibitem{ShorSmolinThapliyal03}
P. W. Shor, J. A. Smolin, and A. V. Thapliyal, 
Phys. Rev. Lett. {\bf 90}, 107901
(2003).
\bibitem{Dur01}
W. D\"ur, Phys. Rev. Lett. {\bf 87}, 230402 (2001).
\bibitem{VedralPlenio98}
V. Vedral and M. B. Plenio, Phys. Rev. A {\bf 57}, 1619 (1998).  
\bibitem{WeiErricsonGoldbartMunro}
T.-C. Wei, M. Ericsson, P. M. Goldbart, and W. J. Munro, Quantum Inf. Comput. {\bf 4}, 252 (2004); eprint quant-ph/0405002.
\bibitem {Kwek02}
D. Kaszlikowski, L. C. Kwek, J. Chen, and C. h. Oh, 
Phys. Rev. A {\bf 66}, 052309 (2002).
\bibitem{SenSenZukowski02}
A. Sen(De), U. Sen, and M. \.Zukowski, 
Phys. Rev. A {\bf 66}, 062318 (2002).
\bibitem{footnote}
In arriving at this result we have made a hypothesis about the
form of the closest separable pure state; see App. A. The result has been verified numerically for the $N=4$ case. 
\bibitem{Horodecki398a}
P. Horodecki, M. Horodecki, and R. Horodecki, 
 Acta Phys. Slov. {\bf 48}, 144 (1998). 
\bibitem{BennettDiVincenzoMorShorSmolinTerhal99}
C. H. Bennett, D. P. DiVincenzo, T. Mor, P. W. Shor, J. A. Smolin, and B. M. Terhal,
Phys. Rev. Lett. {\bf 82}, 5385 (1999). 
\bibitem{Acin02}
A. Ac\'\i n, Phys. Rev. Lett. {\bf 88}, 027901 (2002).
\bibitem{DurCirac00}
W. D\"ur and J. I. Cirac, Phys. Rev. A {\bf 62}, 022302 (2000).
\bibitem{DurCirac00B}
 W. D\"ur and J. I. Cirac, Phys. Rev. A {\bf 61}, 042314 (2000).
\end{thebibliography}
\end{document}